\documentclass[12pt]{JHEP3}
\pdfoutput=1


\usepackage{graphicx}
\usepackage[numbers,sort&compress]{natbib}
\usepackage{amsmath}
\usepackage{amssymb}
\usepackage{marginnote}

\def\beq{\begin{equation}}
\def\eeq{\end{equation}}
\def\beqa{\begin{eqnarray}}
\def\eeqa{\end{eqnarray}}

\makeatletter
\renewcommand\@ENVwarn[1]{}
\makeatother

\title{Holographic deconfinement transition in the presence of a magnetic field}

\author{Alfonso Ballon-Bayona \\
ICTP South American Institute for Fundamental Research, \\
Instituto de  F\'isica Te\'orica, Universidade Estadual Paulista, \\
01140-070 S\~ ao Paulo, SP, Brazil. \\
\email{aballonb@ift.unesp.br} }

\date{today}
\preprint{ICTP-SAIFR/2013-009}

\abstract{We investigate the thermodynamics of the flavour sector of the  Sakai-Sugimoto model  in the presence of a magnetic field. Renormalizing the euclidean Dirac-Born-Infeld action in the antipodal limit, we find for the confined and deconfined phases positive contributions to the pressures that grow with the magnetic field. We also obtain positive magnetic susceptibilities indicating a paramagnetic behaviour of quarks. Using these results we estimate a $\lambda^2/N_c$ correction to the deconfinement temperature that decreases as a function of the magnetic field.}

\begin{document}

\section{Introduction}

A strong magnetic field may be able to change dramatically the QCD phase diagram. It has recently been found in lattice QCD \cite{Bali:2011qj}, from the analyisis of different thermodynamic observables, that the crossover temperature region should fall as the magnetic field increases \cite{Bali:2011qj}. This implies that a strong magnetic field would tend to inhibit confinement. This result has been reproduced qualitatively in \cite{Fraga:2012fs} using the MIT bag model, signalizing the importance of quark confinement. A large-$N_c$ QCD approach considered in \cite{Fraga:2012ev} has also obtained this effect from the analysis of $N_f/N_c$ corrections to the pressure due to quarks degrees of freedom \footnote{A magnetic inhibition of chiral symmetry breaking has also been proposed in \cite{Fukushima:2012kc}.}.

Motivated by these results we investigate the thermodynamics of the quark sector in the Sakai-Sugimoto model \cite{Sakai:2004cn} in the presence of a magnetic field. The Sakai-Sugimoto model is a holographic model for massless QCD in the limit of large-$N_c$. The model describes the confinement of gluons and quarks in the quenched approximation $N_f \ll N_c$. The color sector is described by a D4-brane background in Type IIA String Theory  obtained from $N_c$ coincident D4-branes. One of the spatial coordinates is compactified on a circle wih anti-periodic boundary conditions for the adjoint fermions such that supersymmetry is completely broken.  The quark sector is described by $N_f$ pairs of  $\overline{{\rm D}8}$-D8 branes in the probe limit. At zero temperature the Sakai-Sugimoto model describes confinement and spontaneous chiral symmetry breaking.

An elegant description of deconfinement transition and chiral symmetry restoration in the Sakai-Sugimoto model was done by Aharony, Sonnenschein and Yankielowicz \cite{Aharony:2006da}. At finite temperature a new D4-brane background arises describing a deconfined phase of gluons. The confinement/deconfinement transition then can be mapped to a Hawking-Page transition between two different D4-brane backgrounds. From the analysis of the pressure in both backgrounds a deconfinement temperature was estimated. The result was that gluons should suffer a confinement/deconfinement transition at $T_c = M_{KK}/(2 \pi)$ where $M_{KK}$ is a mass scale of the order of the lightest glueball mass.

In this paper we first review the D4-brane backgrounds (gluon sector) and the $\overline{{\rm D}8}$-D8 brane solutions (quark sector) arising from the euclidean Dirac-Born-Infeld (DBI) action in the presence of an homogeneous magnetic field. These solutions were previously obtained in \cite{Bergman:2008sg,Johnson:2008vna} where chiral symmetry restoration was investigated\footnote{See \cite{Callebaut:2013ria} for a recent study of the two-flavour case.}. For simplicity, we investigate the flavour thermodynamics in the antipodal limit of these configurations where the $\overline{{\rm D}8}$-D8 branes are located at antipodal points of the circle in the fifth dimensional coordinate. The $N_f$ flavour $\overline{{\rm D}8}$-D8 branes can be considered as four-dimensional defects in the 5-d Yang-Mills theory associated with the $N_c$ color D4-branes. As shown in \cite{Aharony:2006da}, in the antipodal limit chiral symmetry restoration and deconfinement occur at the same temperature $T_c = M_{KK}/(2 \pi)$.

We find in this paper, after renormalizing the euclidean DBI action for the $\overline{{\rm D}8}$-D8 branes, a positive contribution to the pressure  for both confined and deconfined phases. These pressures are proportional to $\lambda^3 N_c$ where $\lambda$ is the 't Hooft constant and grow with the magnetic field. The main difference between the pressures is the temperature dependence. In the confined phase the pressure is independent of the temperature suggesting a vacuum interpretation whereas the deconfined phase has a non-trivial dependence on the temperature indicating a plasma interpretation.
From the analysis of the pressure we  estimate a $\lambda^2/N_c$ correction to the deconfinement temperature that is positive at zero magnetic field and becomes negative for strong magnetic fields. The latter is in agreement with the lattice results obtained in \cite{Bali:2011qj}.

\section{Gluons in the Sakai-Sugimoto model}

\subsection{The confined phase}

The confined gluon sector of the Sakai-Sugimoto model is the Witten's model \cite{Witten:1998zw} and consists of the strong coupling description of a stack of $N_c$ D4-branes in Type IIA String Theory. In this model the dual of a 5-d Yang-Mills theory in the large $N_c$ limit is described by a D4-brane background given by
\beqa
ds^2 &=& \left(\frac{u}{R}\right)^{3/2} \left [  dt^2 + d \overline{x}^2 + f(u) d \tau^2  \right ]
+ \left(\frac{R}{u}\right)^{3/2} \left [ \frac{du^2}{f(u)}
+ u^2 d \Omega_4^2 \right ] \, , \cr
e^{\phi}  &=& g_s  \left(\frac{u}{R}\right)^{3/4} \quad , \quad
F_4 = \frac{2 \pi N_c}{V_{S^4}} \epsilon_4 \quad , \quad
f(u) = 1  - \left ( \frac{u_{KK}}{u} \right )^3 \,. \label{confinedbackground}
\eeqa
The D4-brane parameter $R$ is given by
\beqa
R^3 = \pi g_s N_c \, \alpha'^{3/2} \,.
\eeqa
The coordinates $t$ and $\vec{x}$ represent the euclidean time and three spatial coordinates. The extra coordinate $\tau$ is compactified on a circle in order to get a 4-d Yang Mills theory.
Antiperiodic boundary conditions for the adjoint fermions leads to supersymmetry breaking of the effective 4-d Yang-Mills theory. The period in $\tau$ is associated with a 4-d mass scale $M_{KK}$ by $\delta \tau = 2 \pi /(M_{KK})$. The mass scale $M_{KK}$ can be identified with the mass of the lightest glueball.
.

The condition of smoothness of the metric relates $u_{KK}$ with $M_{KK}$ by
\beqa
\frac{2 \pi}{M_{KK}} = \delta \tau = \frac{4 \pi}{3} \frac{R^{3/2}}{u_{KK}^{1/2}} \, .
\eeqa
The euclidean time coordinate $t$  has a period $\beta = 1/T$ where $T$ is identified with the temperature of the dual field theory.

The string coupling is related to the 4-d Yang-Mills coupling through the relation
\beqa
g_s = \frac{ g_{YM}^2 }{ 2 \pi M_{KK} \sqrt{ \alpha'} } \,. \label{gs}
\eeqa
As a consequence of (\ref{gs}), we can write the parameter $R$ in terms of the 't Hooft coupling $\lambda = g_{YM}^2 N_c$ :
\beqa
R^3 = \frac{\lambda}{2 M_{KK}} \alpha' \,.
\eeqa
Note that the 5-d Yang-Mills coupling of the D4-branes, given by
\beqa
g_5^2 = g_{YM}^2 \left ( \frac{4 \pi}{M_{KK}} \right )
= 8 \pi^2 g_s \sqrt{\alpha'} \,,
\eeqa
is a dimensionful quantity. As a consequence there is a power-law running of a dimensionless effective coupling
\beqa
\lambda_5^{eff}(U) = g_5^2 N_c  U
= 4 \pi \frac{ \lambda} {M_{KK}} U \,, \label{powerlaw}
\eeqa
where $U = u/\alpha'$ is the energy of the 5-d theory.

\subsection{The deconfined phase}

As described in \cite{Aharony:2006da}, the deconfined phase of the gluon sector in the Sakai-Sugimoto model is described by the euclidean black brane background given by
\beqa
ds^2 &=& \left(\frac{u}{R}\right)^{3/2} \left [  h(u) dt^2 + d \overline{x}^2 +  d \tau^2  \right ]
+ \left(\frac{R}{u}\right)^{3/2} \left [ \frac{du^2}{h(u)}
+ u^2 d \Omega_4^2 \right ] \, , \cr
e^{\phi}  &=& g_s  \left(\frac{u}{R}\right)^{3/4} \quad , \quad
F_4 = \frac{2 \pi N_c}{V_{S^4}} \epsilon_4 \quad , \quad
h(u) = 1  - \left ( \frac{u_T}{u} \right )^3 \,. \label{deconfinedbackground}
\eeqa

The temperature is again obtained from the euclidean time period as  $\delta t = 1/T$ while the mass scale $M_{KK}$ corresponds to the period of the extra coordinate $\delta \tau = 2 \pi /(M_{KK})$. The condition of smoothness of the geometry now relates the horizon $u_T$ with the temperature :
\beqa
\frac{1}{T} = \delta t = \frac{4 \pi}{3} \frac{R^{3/2}}{u_T^{1/2}} \, .
\eeqa

\subsection{Gluon thermodynamics}

Evaluating the Type IIA Supergravity action for the  D4-brane solutions described in (\ref{confinedbackground}) and (\ref{deconfinedbackground}) and taking into account the Gibbons-Hawking surface terms, it is possible to estimate the finite temperature pressure associated with the dual field theory \cite{Aharony:2006da,Mateos:2007vn}. However, the on-shell action is divergent and have to be regularized such that the divergences can be subtracted. This is done following a holographic renormalization scheme for Dp-brane backgrounds were boundary counterterms are constructed \cite{de Haro:2000xn,Mateos:2007vn,Kanitscheider:2008kd}.

The renormalized actions for the confined and deconfined phases take the form \cite{Aharony:2006da,Mateos:2007vn} :
\beqa
S^{\rm Sugra}_{\rm Conf} &=&  - \left ( \frac{2}{3^7 \pi^2}  M_{KK}^4 \lambda N_c^2 \right) \frac{V_3}{T}  \,, \cr
S^{\rm Sugra}_{\rm Deconf} &=& - \left ( \frac{2}{3^7 \pi^2}  M_{KK}^4 \lambda N_c^2  \right) \frac{V_3}{T} \left ( \frac{2 \pi T}{M_{KK}} \right )^6 \,.
\eeqa
where $V_3$ is the volume of the 3 spatial directions.
Using the thermodynamic relation $P = - T (\partial S/\partial V_3)$, we obtain the corresponding pressures
\beqa
P^{\rm Gluons}_{\rm Conf} &=&  \frac{2}{3^7 \pi^2}  M_{KK}^4 \lambda N_c^2   \, ,
\label{ConfYMPressure} \\
P^{\rm Gluons}_{\rm Deconf} &=& \left ( \frac{2}{3^7 \pi^2}  M_{KK}^4 \lambda N_c^2  \right)
\left ( \frac{2 \pi T}{M_{KK}} \right )^6 \label{DeConfYMPressure}\, . 
\eeqa
The pressure (\ref{ConfYMPressure}) is the vacuum energy arising from gluon diagrams whereas the deconfined result in (\ref{DeConfYMPressure}) can be interpreted in terms of a gluon plasma in the dual field theory. The dependence $\lambda \, T^6$ in (\ref{DeConfYMPressure}) is a consequence of the power-law running of the five dimensional effective coupling (\ref{powerlaw}) in the sense that
\beqa
\lambda \, T^6 \sim \lambda_5 (T) T^5 \, ,
\eeqa
where $\lambda_5 (T) \sim \lambda \, T$ is the effective five dimensional coupling. The $T^5$ dependence  is the expected behaviour in the Stefan-Boltzmann limit of a five dimensional gluon plasma \footnote{Interestingly, a new supergravity background has recently been proposed in \cite{Mandal:2011ws} to describe a four dimensional Yang-Mills plasma  . In this paper, however, we will only consider the  construction of \cite{Aharony:2006da} where the gluon plasma is five dimensional.}. From the condition of equality of pressures  (\ref{ConfYMPressure}) and (\ref{DeConfYMPressure}) one finds in the large-$N_c$ limit a confinement/deconfinement transition at $T_c = M_{KK}/(2 \pi)$.

\section{Quarks in the Sakai-Sugimoto model in the presence of a magnetic field}

As described in the introduction, quarks degrees of freedom in the quenched limit ($N_f \ll N_c$) are described by $N_f$ pair of D8-$\overline{{\rm D}8}$ probe branes living in the D4-brane background. For simplicity we will consider the case of $N_f=1$.

The dynamics of probe branes is described by the Dirac-Born-Infeld (DBI) action. The euclidean DBI action for a D8 brane (or/and $\overline{{\rm D}8}$ brane) can be written as
\beqa
S^{\rm DBI} = \mu_8 \int d^9 x e^{- \phi} \sqrt{ \det (G_{MN} + 2 \pi \alpha' F_{MN})} \, ,
\eeqa
where $G_{MN}$ is the nine dimensional induced metric, $F_{MN}$ is the Maxwell field and $\mu_8 = (2 \pi)^{-8} \alpha'^{-9/2}$ is the D8-brane tension.

The $\overline{{\rm D}8}$-D8 branes define a world volume including the four dimensional coordinates $(t,\vec{x})$ and the sphere angular coordinates $x_\alpha$. The extra coordinate in the world volume can be either $\tau$ or $u$. If we choose $\tau$ then the branes are localized through a curve $u(\tau)$ that satisfies the DBI equations.

The boundary conditions of the problem are the following : at $u \to \infty$ the D8 brane and $\overline{{\rm D}8}$ branes are localized at $\tau = L/2$ and $\tau=-L/2$ respectively whereas in the infrared they may be able to join (or not) at some point $u_0$ if chiral symmetry is broken (preserved). Although we will describe the general solutions for $0 \le L \le \pi/(M_{KK})$, we are only interested in the antipodal solution $L = \pi/(M_{KK})$. The reason is that in the antipodal limit the solution simplifies drastically and it is possible to evaluate the action analytically.

We will turn an homogeneous magnetic field $\vec{B}$ on the $\overline{{\rm D}8}$-D8 branes by considering the following  gauge field ansatz :
\beqa
A_t = A_\tau = A_\alpha = 0 \quad , \quad
\vec{A} = \frac12 \vec{B} \times \vec{x} \, , \label{gaugeansatz}
\eeqa
where $A_\alpha$ denotes the components of the gauge field in the directions of the $S^4$ sphere.

\subsection{The confined phase}

In the confined phase, the induced metric of the D8-$\overline{{\rm D}8}$ branes can be written as
\beqa
ds^2_{D8} = \left ( \frac{u}{R} \right )^{3/2} \Big \{ dt^2
+ d \overline{x}^2 +  \left [ f(u)
+ \left ( \frac{R}{u} \right )^3 \frac{u'^2}{ f(u)}  \right ] d \tau^2 \Big \}  + R^{3/2} u^{1/2} d \Omega_4^2 \,, \label{inducedmetricConf}
\eeqa
where $u' := du/d\tau$.

Under (\ref{gaugeansatz}) and (\ref{inducedmetricConf}),  the DBI action reduces to
\beqa
S_{\rm Conf}^{\rm DBI} = C \frac{V_3}{T} \int d\tau \, u^4 \sqrt{ f(u)
+ \left ( \frac{R}{u} \right )^3 \frac{u'^2}{ f(u)} }
\sqrt{  1 + \left( \frac{R}{u} \right)^3 (2 \pi \alpha')^2  |\vec{B}|^2 }  \, ,
\eeqa
where
\beqa
C := \frac{\mu_8}{g_s} V_{S^4}
= (2 \pi)^{-7} M_{KK}  V_{S^4}  \, \lambda^{-1} \alpha'^{-4} \, N_c \,.
\eeqa
Defining the dimensionless variables $v = u/u_{KK}$ and $\theta = M_{kk} \tau$, the action can be written as
\beqa
S_{\rm Conf}^{\rm DBI} =  \hat C  \frac{V_3}{T} \int_{-\pi}^{\pi}  d \theta \, v^4
\sqrt{ \hat f(v) + \frac94 \frac{{\dot v}^2}{ v^3 \hat f(v)} }
\sqrt{ 1 + \frac{\hat B^2}{4 v^3 } }  \,, \label{D8simplelow}
\eeqa
where
\beqa
\hat C &:=& C u_{KK}^4 M_{KK}^{-1} = \frac{1}{\pi^2} M_{KK}^4 \, \bar \lambda^3 \, N_c \, , \cr
\hat f(v) &:=& 1 - v^{-3} \quad , \quad
\dot v := dv /d\theta \, , \cr
\hat B &:=&  |\vec{B}|/(\bar \lambda M_{KK}^2)  \quad , \quad
\bar \lambda := \lambda/(27 \pi) \, ,
\eeqa
and we used $V_{S^4} = (8/3) \pi^2$. Working in these coordinates it is clear that the $\overline{{\rm D}8}$-D8 brane action is of order $\lambda^3 N_c$.

Since the Lagrangian in (\ref{D8simplelow}) does not depend explicitly on $\theta$, it is useful to consider $\theta$ as a time direction so that the Euler-Lagrange equations are equivalent to the conservation of the Hamiltonian
\beqa
\frac{d H}{d \theta} = - \hat C \frac{V_3}{T} \frac{ d }{d \theta} \Big \{ v^4 \hat f(v)
\frac { \sqrt{ 1 + \frac{\hat B^2}{4 v^3 } } }
{ \sqrt{ \hat f(v) + \frac94 \frac{{\dot v}^2}{ v^3 \hat f(v)} } } \Big \} = 0 \,.  \label{constantHv}
\eeqa
The boundary conditions  in the ultraviolet are $v(\pm \Delta \theta/2) = \infty$, where $\Delta \theta  = M_{KK} L$ where $\pm \Delta \theta/2$ are the positions of the D8 and $\overline{{\rm D}8}$ in the $\theta$ coordinate. At the minimum value $v=1$   (corresponding to $u=u_{KK}$) the size is effectively zero. For this reason the D8 and $\overline{{\rm D}8}$ branes have to connect at some point $ v_0 \ge 1$ and chiral symmetry is spontaneously broken.
Since the solution is even in $\theta$ we have the infrared boundary conditions $v(0) = v_0$ and $\dot v(0) = 0$. From the Hamiltonian conservation (\ref{constantHv}) and the infrared boundary conditions we find
\beqa
v^4 \hat f(v)
\frac { \sqrt{ 1 + \frac{\hat B^2}{4 v^3 } } }
{ \sqrt{ \hat f(v) + \frac94 \frac{{\dot v}^2}{ v^3 \hat f(v)} } }
= v_0^4 \sqrt{ \hat f(v_0)} \sqrt{ 1 + \frac{\hat B^2}{4 v_0^3 } } \, . \eeqa
Solving this equation we find
\beqa
\frac{dv}{d \theta} = \pm \frac23  v^{3/2} \hat f(v) \, \sqrt{
\frac{v^8}{v_0^8} \frac{ \hat f(v)}{ \hat f(v_0) }
\frac{\left ( 1 + \frac{\hat B^2}{4 v^3 } \right ) }{ \left ( 1 + \frac{\hat B^2}{4 v_0^3 } \right ) } - 1 }  \,, \label{soldotv}
\eeqa
where $\pm$ corresponds to the D8 ($\overline{{\rm D}8}$) sectors.
Evaluating (\ref{D8simplelow}) at the solution (\ref{soldotv}), we find that the on-shell action takes the form
\beqa
S_{\rm Conf}^{\rm DBI} = 3 \hat C \frac{V_3}{T} \int_{v_0}^{\infty} dv \,
\frac{v^{5/2} \left ( 1 + \frac{ \hat B^2}{4 v^3} \right ) }
{ \sqrt { \hat f(v) \left ( 1 + \frac{ \hat B^2}{4 v^3} \right )
- \frac{v_0^8}{v^8} \hat f(v_0) \left ( 1 + \frac{ \hat B^2}{4 v_0^3} \right ) } } \,.
\eeqa

The antipodal limit corresponds to $v_0 \to 1$ (or $u_0 \to u_{KK}$). In that limit  we find $\dot v \to \pm \infty$ and the D8-$\overline{{\rm D}8}$ branes become two horizontal lines at $\theta = \pm \pi/2$ that are smoothly connected at $u_{KK}$. In the antipodal limit the on-shell action reduces to
\beqa
S_{\rm Conf}^{\rm DBI} =  3 \hat C  \frac{V_3}{T} \int_1^{\infty} d v \,
\frac{v^{5/2}}{\sqrt{ \hat f(v)}} \sqrt{ 1 + \frac{\hat B^2}{4 v^3 } }  \,. \label{D8lowAP}
\eeqa

\subsection{The deconfined phase}

In this case the induced metric takes the form
\beqa
ds^2_{D8} = \left ( \frac{u}{R} \right )^{3/2} \Big \{ h(u) dt^2
+ d \overline{x}^2 +  \left [ h(u)
+ \left ( \frac{R}{u} \right )^3 u'^2  \right ] \frac{d \tau^2}{h(u)} \Big \}  + R^{3/2} u^{1/2} d \Omega_4^2 \,. \label{inducedmetricDeconf}
\eeqa

Under the gauge field ansatz (\ref{gaugeansatz}) the DBI action takes the form
\beqa
S_{\rm Deconf}^{\rm DBI} =  \hat C  \frac{V_3}{T} \int d \theta \, v^4
\sqrt{ \hat h(v) + \frac94 \frac{{\dot v}^2}{ v^3 } }
\sqrt{ 1 + \frac{\hat B^2}{4 v^3 } }  \,, \label{D8simpleHighT}
\eeqa
where $\hat h(v) := 1 - (v_T/v)^3$ with $v_T = u_T/u_{KK}$.
The conservation of Hamiltonian associated with (\ref{D8simpleHighT}) now reads
\beqa
 \frac{ d H}{d \theta} = - \hat C \frac{V_3}{T} \frac{d}{d \theta} \Big \{ v^4 \hat h(v)
\frac{\sqrt{ 1 + \frac{\hat B^2}{4 v^3 } }}
{\sqrt{ \hat h(v) + \frac94 \frac{{\dot v}^2}{ v^3 } } } \Big \} = 0 \,.
\eeqa

Since the $\tau$ coordinate does not shrink to zero size at the horizon $v_t$, now we have two possible solutions. The first solution describes a scenario where the $D8$ and $\overline{{\rm D}8}$ branes merge at some point $v_0$ and chiral symmetry is broken. This solution takes the form
\beqa
\dot v = \pm \frac23  v^{3/2} \sqrt{ \hat h(v)}
\sqrt{ \frac{v^8}{v_0^8} \frac{ \hat h(v)}{\hat h(v_0)}
\frac{ \left (1 + \frac{ \hat B^2}{4 v^3} \right ) }
{\left (1 + \frac{ \hat B^2}{4 v_0^3} \right ) } - 1 } \,.
\eeqa

The corresponding on-shell action takes the form
\beqa
S_{\rm Deconf}^{\rm DBI (I)} = 3 \hat C \frac{V_3}{T} \int_{v_0}^{\infty} dv \,
\frac{v^{5/2} \sqrt{ \hat h(v)} \left ( 1 + \frac{ \hat B^2}{4 v^3} \right ) }
{ \sqrt { \hat h(v) \left ( 1 + \frac{ \hat B^2}{4 v^3} \right )
- \frac{v_0^8}{v^8} \hat h(v_0) \left ( 1 + \frac{ \hat B^2}{4 v_0^3} \right ) } } \,.
\eeqa

The second solution describes a scenario where the D8 and $\overline{{\rm D}8}$ branes are disconnected. They remain localized at $\theta = \Delta \theta/2$ and $\theta = - \Delta \theta/2$ all the way ending separately at $v_T$ so that chiral symmetry is preserved. The on-shell action then takes the form
\beqa
S_{\rm Deconf}^{\rm DBI (II)} &=& 3 \hat C \frac{V_3}{T} \int_{v_T}^{\infty} d v \, v^{5/2}
\sqrt{  1 +  \frac{ \hat B^2}{4 v^3} } \label{D8highAP}
\, .
\eeqa

In the antipodal limit $\Delta \theta \to \pi$ and the first solution disappears. This implies that chiral symmetry restoration is mandatory in the deconfined phase. The on-shell action in the antipodal limit is given by (\ref{D8highAP}).

\subsection{Quark thermodynamics}

\subsubsection{Maxwell truncation}

The effective magnetic field $\hat B$ goes to zero in the limit of large $\lambda$,  keeping $|\vec{B}|$ fixed. Then  we can expand the square root in (\ref{D8lowAP}) and (\ref{D8highAP}) in powers of $\hat B^2$. The result for the confined phase is
\beqa
S_{\rm Conf}^{\rm DBI} &=&  3 \hat C  \frac{V_3}{T} \Big [ \int_1^{\infty} d v \,
\frac{v^{5/2}}{\sqrt{ \hat f(v)}}
+ \frac{ \hat B^2}{8 } \int_1^{\infty} d v \,
\frac{v^{-1/2}}{\sqrt{ \hat f(v)}} + {\cal O} (\hat B^4) \Big ] \cr
&=&  3 \hat C  \frac{V_3}{T} \Big [ I_{-7/6,1/2}
+ \frac{ \hat B^2}{8 }  I_{-1/6,1/2} + {\cal O} (\hat B^4) \Big ] \,, \label{D8lowAPExpanded}
\eeqa
where
\beqa
I_{a,b} :=  \frac13 \int_0^1 dx \, x^{a-1} (1-x)^{b-1}   \, ,
\eeqa
and we performed the change of variable $x=v^{-3}$.  The integral $I_{a,b}$ is finite when $a>0$ and $b>0$. However in our case the integrals are divergent so they need to be regularized. Introducing a cut-off at $v_{max}=1/\epsilon$ (equivalently $x_{min}= \epsilon^3$) with $\epsilon \to 0$, the integrals can be performed and we obtain
\beqa
S_{\rm Conf}^{\rm DBI}
&=& \hat C \frac{V_3}{T} \Big [ \frac67 \epsilon^{-7/2} + 3 \epsilon^{-1/2}  + \frac34 \hat B^2 \epsilon^{-1/2} \cr
&+& B(-7/6,1/2) + \frac{ \hat B^2}{8 }  B(-1/6,1/2) + {\cal O} (\hat B^4) \Big ] \, ,
\label{divergentconf}
\eeqa
where  $B(a,b)$ is the Beta function, which is usually written as
\beqa
B(a,b) = \frac{ \Gamma(a) \Gamma(b)}{\Gamma (a+b)} \,.
\eeqa
The divergences in (\ref{divergentconf}) can be cancelled out with a counterterm action of the form
\beqa
S_{\rm ct} &=&  - \hat C \frac{V_3}{T}  \Big [  \frac67 \epsilon^{-7/2} + 3 ( 1 +  \frac{\hat B^2}{4} ) \epsilon^{-1/2}  \Big ]\,.  \label{countertermconf}
\eeqa
This counterterm can be obtained by considering the variational problem of the DBI action (\ref{D8simplelow}) taking the $v$ coordinate as a field depending on $\theta$. The variation of (\ref{D8simplelow}) takes the form
\beqa
\delta S^{\rm DBI}_{\rm Conf} = \int_{-\pi}^{\pi} d \theta \Big [ \frac{\partial L^{\rm DBI}_{\rm Conf}}{\partial  v} -  \frac{d \Pi^{\rm DBI}_{\rm Conf} }{d \theta} \Big ] \delta v
\,+\, \Big [ \Pi^{\rm DBI}_{\rm Conf} \, \delta v \Big ]_{-\pi}^{\pi} \, ,  \label {deltaSDBIConf}
\eeqa
where 
\beqa
\Pi^{\rm DBI}_{\rm Conf} := \frac{\partial L^{\rm DBI}_{\rm Conf}}{\partial \dot v} \, . 
\eeqa
When evaluating (\ref{deltaSDBIConf}) at the solution (\ref{soldotv}) the bulk term vanishes and the boundary term takes the form 
\beqa
\Big [ \Pi^{\rm DBI}_{\rm Conf} \, \delta v \Big ]_{-\pi}^{\pi} = \hat C \frac{V_3}{T} \,  \delta \left [ \frac67 v_{max}^{7/2} + 3 ( 1 +  \frac{\hat B^2}{4} ) v_{max}^{1/2} 
+ \dots \right ] \,.  \label{deltaSDBIConfBdy}
\eeqa
Thus the variation of the counterterm action (\ref{countertermconf}) cancels the first two terms in  (\ref{deltaSDBIConfBdy}) whereas the subleading terms can be cancelled by adding counterterms of the form  $\epsilon^{n + 1/2}$ with $n$ some positive integer. This way the variational problem is now well defined and we have also cancelled the divergences. The reason why $\delta v$ is not zero in (\ref{deltaSDBIConfBdy}) is that  we have already imposed the infrared boundary conditions  $v(0) = v_0$ and $\dot v(0) = 0$ to get the solution (\ref{soldotv}) so there is no freedom to fix $v$ again. 

We have adopted here a minimal prescription where the counterterm action (\ref{countertermconf}) does not include any unnecessary finite terms. For this minimal substraction the total action takes the form
\beqa
S_{\rm Conf}^{\rm DBI}
&=& \hat C \frac{V_3}{T}
B(-7/6,1/2) \Big [ 1 + \frac{7}{32} \hat B^2 + {\cal O} (\hat B^4) \Big ] \, ,
\label{RegMaxConf}
\eeqa
and we have used the the recurrence relation $(a+b) B(a+1,b)= a B(a,b)$. At $\hat B=0$ the result (\ref{RegMaxConf}) may be interpreted in terms of a (negative) Casimir energy associated with the D8-$\overline{{\rm D}8}$ probe branes.

For the deconfined phase we can expand the square root in the action (\ref{D8highAP}) and get
\beqa
S_{\rm Deconf}^{\rm DBI}  = 3 \hat C \frac{V_3}{T}
\Big [ \int_{v_T}^{\infty} v^{5/2}
+ \frac{ \hat B^2}{8 }
\int_{v_T}^{\infty} dv v^{-1/2} + {\cal O} (\hat B^4) \Big ] \,.
\eeqa
This time the integrals are trivial so that after introducing a cut-off at $v_{max}=1/\epsilon$ we get
\beqa
S_{\rm Deconf}^{\rm DBI}  =  \hat C \frac{V_3}{T}
\Big [ \frac{6}{7} \epsilon^{-7/2}
+ \frac34  \hat B^2  \epsilon^{-1/2}
- \frac{6}{7} v_T^{7/2} - \frac34 \hat B^2 v_T^{1/2}  + {\cal O} (\hat B^4) \Big ] \,.
\label{divergentdeconf}
\eeqa
Note that one divergence present in the confined regime is not present in the deconfined case. This is a consequence of the difference on the actions in both regimes. 
This time the counterterm action takes the form
\beqa
S_{\rm ct} &=&  - \hat C \frac{V_3}{T}  \Big [ \frac67 \epsilon^{-7/2} + \frac34  \hat B^2 \epsilon^{-1/2}  \Big ]\, .   \label{countertermdeconf}
\eeqa
Again, this action can be  obtained by considering the variational problem for the action (\ref{D8simpleHighT}). In the minimal prescription the total action takes the form
\beqa
S_{\rm Deconf}^{\rm DBI} = - \hat C \frac{V_3}{T}
\bar T^7 \left (\frac67 \right) \Big [ 1
+ \frac78 \frac{ \hat B^2 } { \bar T^6 }+ {\cal O} (\hat B^4) \Big ] \, , \label{RegMaxDeconf}
\eeqa
where we have defined  $\bar T :=  2 \pi T/(M_{KK})$ and we used
\beqa
v_T = \frac{u_T}{u_{KK}} = \bar T^2 \, .
\eeqa

We have used a renormalization procedure based on the principle of least action and is similar to the one used for holographic Wilson loops \cite{Drukker:1999zq,Drukker:2005kx} and to the one suggested in \cite{Andrade:2006pg} for one-point functions. It would be interesting to develop a covariant procedure more similar to that used previously in \cite{Mateos:2007vn,Benincasa:2009ze}. The main difficulty for that construction arises due to the non-conformal asymptotics of the Sakai-Sugimoto model. One may need to consider the Weyl transformation  for Dp-brane backgrounds suggested in \cite{Kanitscheider:2008kd,Kanitscheider:2009as,Benincasa:2009ze}, which maps  the metric to an asymptotically $AdS_{p+2} \times S^{8-p}$ spacetime.

\subsubsection{Complete DBI action}

Redefining the variables as $x=v^{-3}$ we can write the action for the confined phase (\ref{D8lowAP}) as
\beqa
S_{\rm Conf}^{\rm DBI} =  \hat C  \frac{V_3}{T}
\int_0^1 dx \, x^{a-1} (1-x)^{b-1} (1 - z x)^{-c} \, ,
\eeqa
where
\beqa
a =  - \frac76 \quad, \quad b = \frac12 \quad , \quad c = - \frac12 \quad , \quad
z = - \frac { \hat B^2}{4} \,.
\eeqa
We regularize the integral and the divergent part can be subtracted using the  counterterm action (\ref{countertermconf}) described in the Maxwell case. The reason is that the terms of order $\hat B^4$ or higher in the expansion do not bring new divergences.
The total action reads
\beqa
S_{\rm Conf}^{\rm DBI} =  \hat C \frac{V_3}{T}
B(-7/6,1/2) \,  F \left ( - \frac12, - \frac76 ; - \frac23 ; - \frac{ \hat B^2}{4} \right) \,, \label{RegDBIConf}
\eeqa
where $F(c,a;a+b;z)$ is the hypergeometric function. Expanding this function for small $z$ we recover the Maxwell result (\ref{RegMaxConf}).

For the deconfined phase we define the variable $x=(v_T/v)^3$ so that the action (\ref{D8highAP}) takes the form
\beqa
S_{\rm Deconf}^{\rm DBI} =  \hat C \frac{V_3}{T} v_T^{7/2} \int_0^1 d x \, x^{a-1} (1 - x)^{b-1} (1 - z x)^{-c}  \,, 
\eeqa
where
\beqa
a = -7/6 \quad , \quad b = 1 \quad , \quad c = -1/2
\quad , \quad z = - \frac{ \hat B^2}{4 v_T^3}\,.
\eeqa
Regularizing the integral and subtracting the divergences with the counterterm action (\ref{countertermdeconf}) we get
\beqa
S_{\rm Deconf}^{\rm DBI} = -  \hat C  \frac{V_3}{T} \bar T^7 \left ( \frac67 \right )  F \left(-\frac12,-\frac76 ;-\frac16 ;- \frac{ \hat B^2}{4 \bar T^6} \right )  \,.
\label{RegDBIDeconf}
\eeqa
Expanding the hypergeometric function for small argument we recover the Maxwell result (\ref{RegMaxDeconf}).

\subsection{Pressures, magnetizations and magnetic susceptibilities}

The quark contribution to the pressure in the confined regime is given by
\beqa
P_{\rm Conf}^{\rm Quarks} =  -  \frac{M_{KK}^4}{\pi^2} \bar \lambda^3 N_c   B(-7/6,1/2)   \,  F  \left( - \frac12, - \frac76 ; - \frac23 ; - \frac{ \hat B^2}{4} \right ) \,. \label{ConfQuarkPressure}
\eeqa
Since $B(-7/6,1/2) \approx  -2.56057$  we conclude that the quark contribution to the pressure is positive in the confined regime. This result may be interpreted as a $\lambda^2/N_c$ correction to the vacuum pressure, arising from quark loops \footnote{This result differs from the result found in \cite{Fraga:2012ev} where a negative contribution to the pressure was found from a $1/N_c$ correction to the effective string tension.}.

The pressure in the deconfined regime takes the form
\beqa
P_{\rm Deconf}^{\rm Quarks} &=&  \frac67   \frac{M_{KK}^4}{\pi^2} \bar \lambda^3 N_c  \, \bar T^7    \, F \left (-\frac12,-\frac76 ;-\frac16 ;- \frac{\hat B^2}{4 \bar T^6} \right )  \,,
\label{DeconfQuarkPressure}
\eeqa
This result can be interpreted as a $\lambda^2/N_c$ correction to the pressure of a strongly coupled quark gluon plasma where the gluons live in five dimensions (the world volume of the D4-branes) and the quarks live in four dimensional defects (the intersection of the D4-branes and D8-$\overline{{\rm D}8}$ branes).

Of course, this plasma is very different from the four dimensional quark gluon plasma developed in real QCD. In particular, the dependence  $\lambda^3 T^7$ at zero magnetic field is a consequence of the power-law running of the effective coupling described in (\ref{powerlaw}) in the sense that
\beqa
\lambda^3 T^7  \sim \lambda_5^3 (T) T^4  \, ,
\eeqa
where $\lambda_5 (T) \sim \lambda T$ is the effective five dimensional coupling. The $T^4$ dependence is what we already expect from the Stefan-Boltzmann limit.

\begin{figure}[t]
\begin{center}
\includegraphics[width=.5\textwidth]{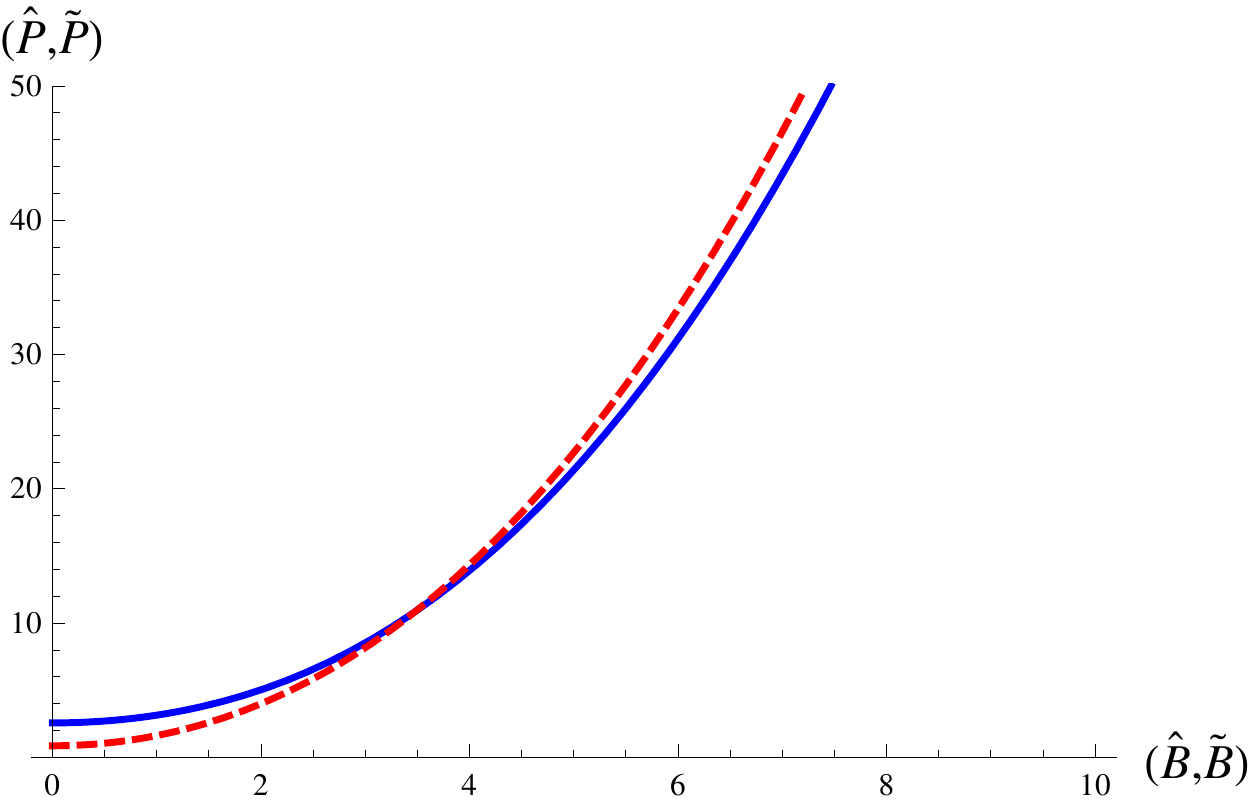}
\caption{\label{Pressure}  The blue solid line represents the rescaled confined pressure $\hat P $ as a function of the rescaled magnetic field $\hat B$. The red dashed line represents the rescaled deconfined pressure $\tilde P$ as a function of  $\tilde B =  \hat B \bar T^{-3}$. Read the text for a description of the rescaled quantities. }
\end{center}
\end{figure}

Although the theory that we are leading with is not real QCD, our result may be useful as a guide towards the physics of a strongly coupled quark gluon plasma in the presence of a magnetic field. 
We plot in Figure \ref{Pressure} our results for the confined and deconfined quark pressures. For the confined phase we define a rescaled pressure $\hat P = P_{\rm Conf}^{\rm Quarks} \times \left( \frac{M_{KK}^4}{\pi^2} \bar \lambda^3 N_c \right )^{-1}$ and plot it as a function of the rescaled magnetic field $\hat B = |\vec{B}| \times ({\bar \lambda} M_{KK}^2)^{-1}$.  This is shown in the blue solid line. For the deconfined phase we find convenient to define the rescaled pressure $\tilde P = P_{\rm Deconf}^{\rm Quarks} \times \left( \frac{M_{KK}^4}{\pi^2} \bar \lambda^3 N_c \right )^{-1} \bar T^{-7} $ and plot it as a function of a combination of a new variable $\tilde B =  \hat B \bar T^{-3}$. From the plots we can conclude that, at non-zero magnetic field and fixed temperature, the magnetization is positive in the confined and deconfined regime. Interestingly, if we consider $T_c = M_{KK}/(2 \pi)$ we find that the difference of pressures is non-zero and  changes sign at $\hat B \approx 3.48$.

\begin{figure}[t]
\includegraphics[width=0.45\textwidth]{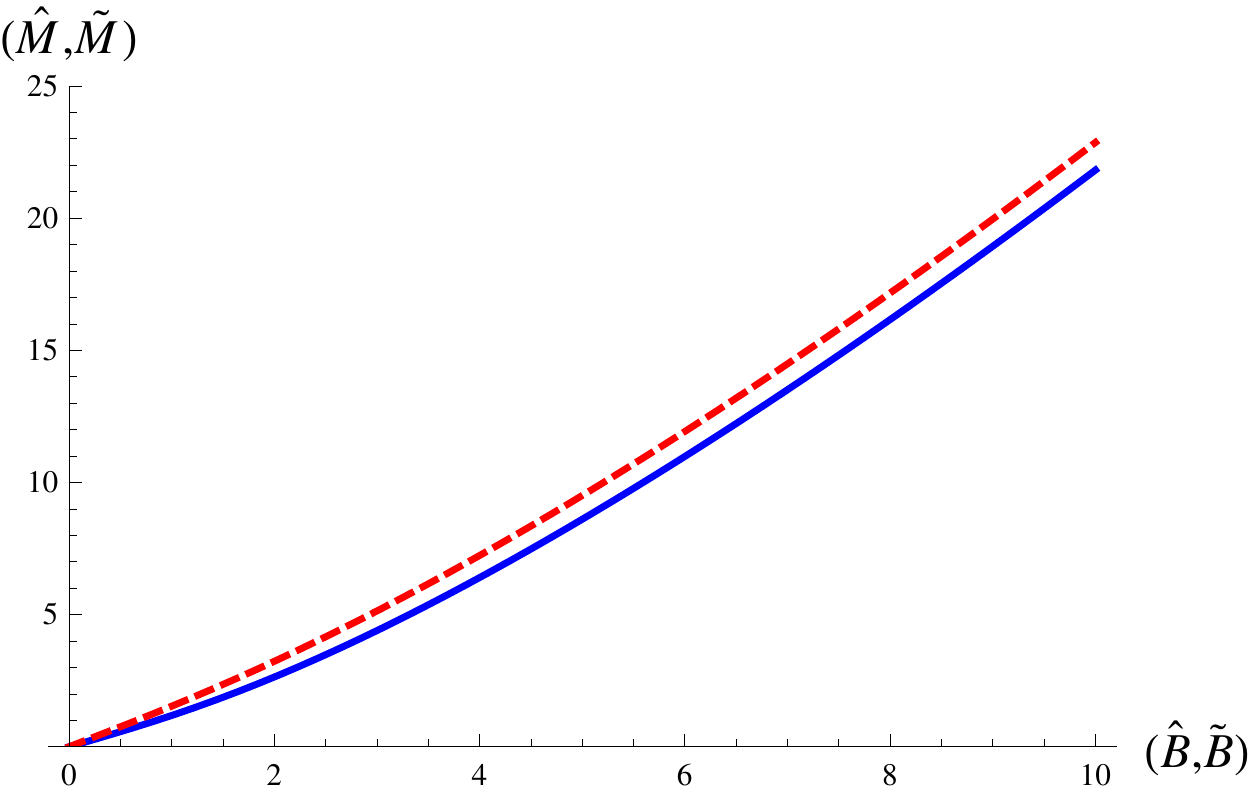}
\hspace{1cm}
\includegraphics[width=0.45\textwidth]{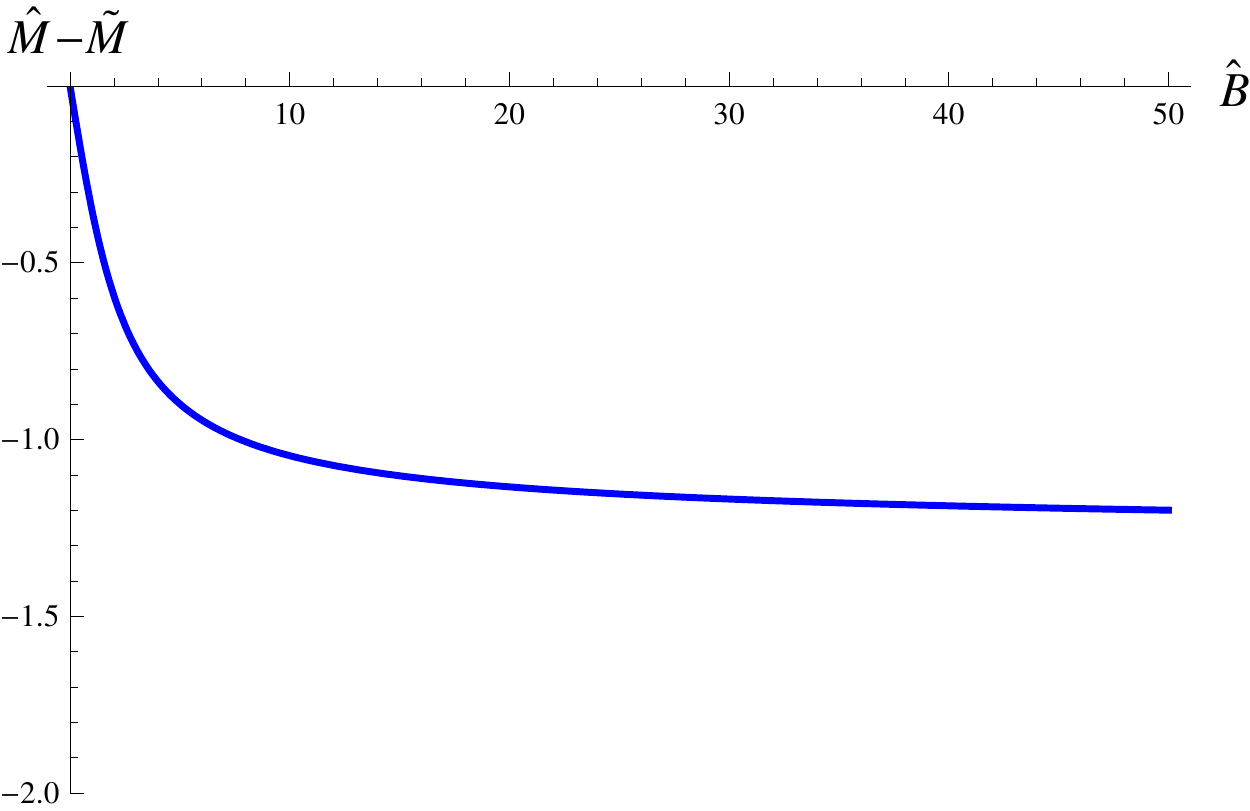}
\caption{\label{Magnetization}  Left Panel: The blue solid line represents the rescaled confined magnetization $\hat M $ as a function of the rescaled magnetic field $\hat B$. The red dashed line represents the rescaled deconfined magnetization $\tilde M$ as a function of  $\tilde B =  \hat B \bar T^{-3}$ (read the text for a description of the rescaled quantities). Right Panel : The magnetization difference $\hat M - \tilde M$ at $\bar T =1$ as a function of the rescaled magnetic field $\hat B$.  }
\end{figure}

The magnetizations and magnetic susceptibilities are determined from the euclidean DBI actions through the relation
\beqa
M (B,T) &=& - \frac{T}{V_3} \frac{ \partial S }{\partial |\vec{B}|} 
= - \bar \lambda^{-1} M_{KK}^{-2} \frac{T}{V_3} \frac{ \partial S }{\partial \hat B}  \, , \cr 
\chi(B,T) &=&  \frac{ \partial M }{\partial |\vec{B}|} 
= \bar \lambda^{-1} M_{KK}^{-2}  \frac{ \partial M }{\partial \hat B} \,.
\eeqa

In the confined phase the magnetization and magnetic susceptibility are independent of the temperature and take the form
\beqa
M_{\rm Conf} (\hat B) &=& - \frac{M_{KK}^2}{\pi^2} \bar \lambda^2 N_c \left ( \frac{7}{16} \right ) B(-7/6,1/2) \,  F \left ( \frac12, - \frac16 ; \frac13 ; - \frac{ \hat B^2}{4} \right) \, \hat B \, ,  \label{ConfMagnetization} \\
\chi_{\rm Conf} (\hat B) &=& - \frac{ \bar \lambda N_c}{\pi^2} 
\left ( \frac{7}{16} \right ) B(-7/6,1/2) \Big [  F \left (  \frac12, - \frac16 ;  \frac13 ; - \frac{ \hat B^2}{4} \right) \cr 
&+& \frac{\hat B^2}{8} \,  F \left (  \frac32,  \frac56 ;  \frac43 ; - \frac{ \hat B^2}{4} \right)\Big ]  \,.
\label{ConfSusceptibility}
\eeqa

In the deconfined phase we find the following magnetization and magnetic susceptibility :
\beqa
M_{\rm Deconf}  (  \hat B \bar T^{-3}, \bar T  ) &=& \frac{ M_{KK}^2}{\pi^2} \bar \lambda^2 N_c
\left ( \frac32 \right )   \,  F  \left ( \frac12,-\frac16 ; \frac56 ;- \frac{\hat B^2}{4 \bar T^6} \right ) \, \hat B \, \bar T  \, , 
\label{DeconfMagnetization} \\
\chi_{\rm Deconf} &=&  
\frac{ \bar \lambda N_c}{\pi^2} \left ( \frac32 \right ) \bar T \Big [ F \left( \frac12, - \frac16 ; \frac56 ; - \frac{ \hat B^2}{4 \bar T^6} \right ) \cr 
&\,+\,& \frac{ \hat B^2}{20  \bar T^6} \, F \left ( \frac32 , \frac56 ; \frac{11}{6} ; - \frac{ \hat B^2}{4 \bar T^6} \right )
\Big ]\, . \label{DeconfSusceptibility}
\eeqa

From the results (\ref{ConfMagnetization}) and (\ref{DeconfMagnetization}) it is clear that when the magnetic field is small the dependence of the confined and deconfined magnetizations is linear in $\hat B$. To analyse the opposite regime we rewrite the magnetizations as 
\beqa
M_{\rm Conf} (\hat B) &=& - \frac78 \frac{M_{KK}^2}{\pi^2} \bar \lambda^2 N_c  B(-7/6,1/2) \cr 
&\times&  \Big [ \frac{ \Gamma(1/3) \Gamma(2/3)}{\Gamma^2(1/2)} \left ( \frac{ \hat B}{2}\right )^{4/3} F \left (- \frac16, \frac12, \frac13; - \frac{4}{\hat B^2} \right ) \cr 
&+& \frac{ \Gamma(1/3) \Gamma(-2/3)}{\Gamma^2(-1/6)} F \left (\frac12, \frac76 ; \frac53 ; - \frac{4}{\hat B^2} \right ) \Big ] \, ,  \label{ConfMagnetizationLargeB} \\
M_{\rm Deconf}  (  \hat B \bar T^{-3}, \bar T  ) &=& 3 \frac{ M_{KK}^2}{\pi^2} \bar \lambda^2 N_c  \bar T^4 \cr 
&\times&  \Big [ \frac{\Gamma(5/6) \Gamma(2/3)}{\Gamma(1/2)} 
\left ( \frac{ \hat B}{2 \bar T^3} \right )^{4/3} F \left ( - \frac16,0; \frac13; - \frac{4 \bar T^6}{\hat B^2} \right ) \cr 
&+& \frac{\Gamma(1/2) \Gamma(-2/3)}{\Gamma(-1/6) \Gamma (1/3)} F \left ( \frac12 , \frac23; \frac53;- \frac{4 \bar T^6}{\hat B^2} \right ) \Big ]    \, , 
\label{DeconfMagnetizationLargeB}
\eeqa
From (\ref{ConfMagnetizationLargeB}) and (\ref{DeconfMagnetizationLargeB}) we conclude that the leading order of the magnetization  is $\hat B^{4/3}$ for both confined  and deconfined phases. This result differs with the $B \log B$ dependence obtained in perturbative QCD calculations \cite{Cohen:2008bk} and D3/D7 holographic calculations \cite{Albash:2007bk} for zero bare mass. Interestingly, at $\bar T=1$ the leading terms in (\ref{ConfMagnetizationLargeB}) and (\ref{DeconfMagnetizationLargeB}) are exactly the same so that the magnetization difference is a constant at $\hat B \to \infty$. This saturation phenomena is similar to that obtained in \cite{Johnson:2008vna} for the non-antipodal solutions in the deconfined regime. 

In Figure \ref{Magnetization} we plot our results for the magnetizations. In the confined regime we define the rescaled magnetization  $\hat M = M_{\rm Conf} \times \left( \frac{ M_{KK}^2}{ \pi^2}  \bar \lambda^2 N_c \right )^{-1}$  whereas in the deconfined regime we define  $\tilde M = M_{\rm Deconf} \times \left( \frac{ M_{KK}^2}{ \pi^2} \bar \lambda^2 N_c \right )^{-1} \bar T^{-1} $. The left panel of Figure \ref{Magnetization} shows a plot of $\hat M$  as a function of  $\hat B$ and $\tilde M$ as a function of $\tilde B =  \hat B \bar T^{-3}$. The right panel of Figure \ref{Magnetization} shows the difference of magnetizations $\hat M - \tilde M$ for $\bar T =1$ as a function of the magnetic field $\hat B$, indicating a saturation effect at large $\hat B$. 

\begin{figure}[t]
\begin{center}
\includegraphics[width=.5\textwidth]{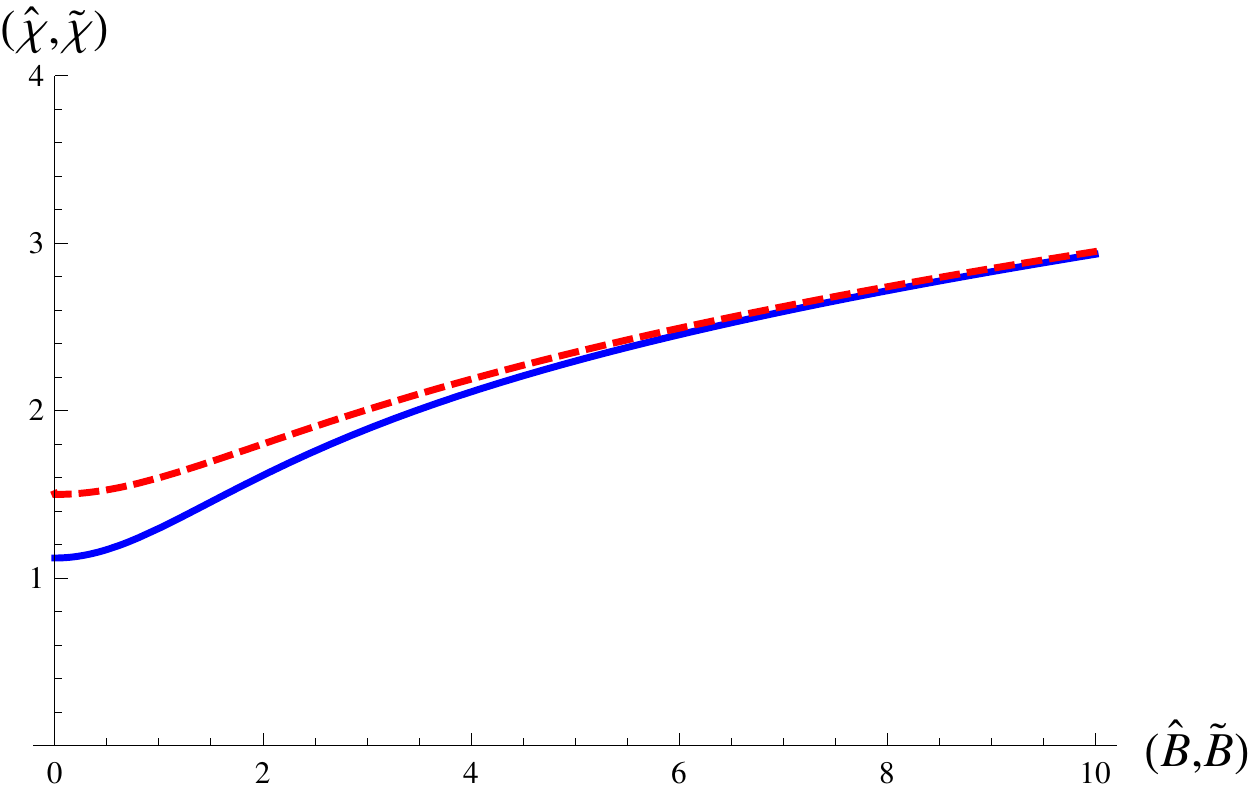}
\caption{\label{Susceptibilities} The blue solid line represents the rescaled confined magnetic suceptibility $\hat \chi$ as a function of the rescaled magnetic field $\hat B$. The red dashed line represents the rescaled deconfined magnetic susceptibility $\tilde \chi$ as a function of  $\tilde B =  \hat B \bar T^{-3}$. Read the text for a description of the rescaled quantities.}
\end{center}
\end{figure}

In Figure \ref{Susceptibilities} we plot our results for the magnetic susceptibilities. In the confined regime we define the rescaled magnetic susceptibility  $\hat \chi = \chi_{\rm Conf} \times \left( \frac{ \bar \lambda N_c }{ \pi^2}  \right )^{-1}$ and plot it as a function of $\hat B$, where as in the deconfined regime we define  $\tilde \chi = \chi_{\rm Deconf} \times  \left( \frac{ \bar \lambda N_c }{\pi^2}  \right )^{-1} \bar T^{-1} $ and plot it as a function of  $\tilde B =  \hat B \bar T^{-3}$. From these plots we conclude that the magnetic susceptibilities increase with the magnetic field. Interestingly, at zero magnetic field we find a positive result for the confined and deconfined magnetic  susceptibility that indicates a {\it paramagnetic response} of quarks in the strong coupling regime. 

\begin{figure}[t]
\begin{center}
\includegraphics[width=0.5\textwidth]{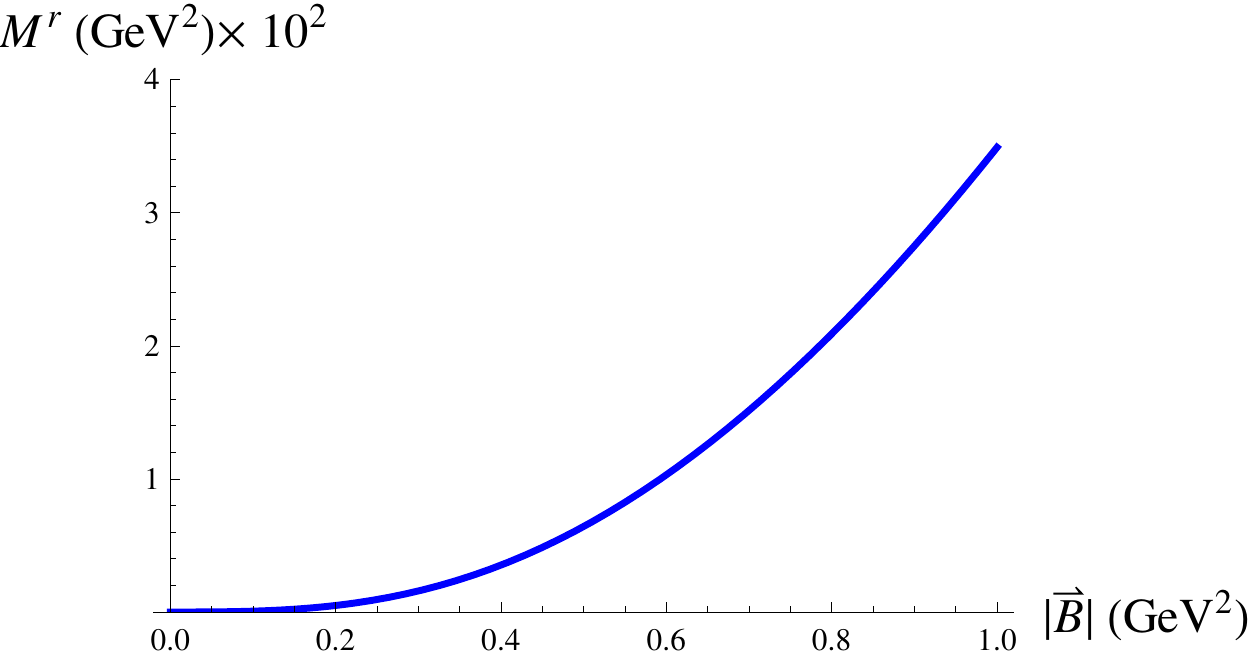}
\caption{\label{RealisticMagnetization} The vacuum magnetization after subtracting the linear term, i.e. $M^r( |\vec{B}|) := M_{\rm Conf}( |\vec{B}|) - M_{\rm Conf}'(0) |\vec{B}|$ for the  phenomenological values $M_{KK}= 1 {\rm GeV}$, $ \bar \lambda = \frac13$ and $N_c=3$.}
\end{center}
\end{figure}

In Figure \ref{RealisticMagnetization} we plot $M^r( |\vec{B}|) := M_{\rm Conf}( |\vec{B}|) - M_{\rm Conf}'(0) |\vec{B}|$ for the  phenomenological values $M_{KK}= 1 {\rm GeV}$, $ \bar \lambda = \frac13$ and $N_c=3$. This shows that the vacuum remains paramagnetic even after subtracting the linear term in the magnetization. A similar paramagnetic behaviour has been predicted  recently using a hadron resonance gas model \cite{Endrodi:2013cs} and observed in a lattice QCD calculation \cite{Bali:2013esa}.

\subsection{Correction to the deconfinement temperature}

As described in section {\bf 2}, the gluon contribution to the pressure is of order $\lambda N_c^2$. A deconfinement temperature $T_c = M_{KK}/(2 \pi)$ was obtained by equating the gluon pressures in the confined and deconfined phases. On the other hand, the quark contribution to the pressure is of order $\lambda^3 N_c$ and we have seen from Figure \ref{Pressure} that at $T_c = M_{KK}/(2 \pi)$ the result for the confined phase differs from that obtained in the deconfined phase. This means that the total pressures evaluated at $T_c$ are not equal :
\beqa
P_{\rm Conf}^{\rm Gluons}(T_c) +  P_{\rm Conf}^{\rm Quarks}(T_c)  \ne
P_{\rm Deconf}^{\rm Gluons}(T_c) +  P_{\rm Deconf}^{\rm Quarks} (T_c)
\eeqa

\begin{figure}[t]
\begin{center}
\includegraphics[width=.5\textwidth]{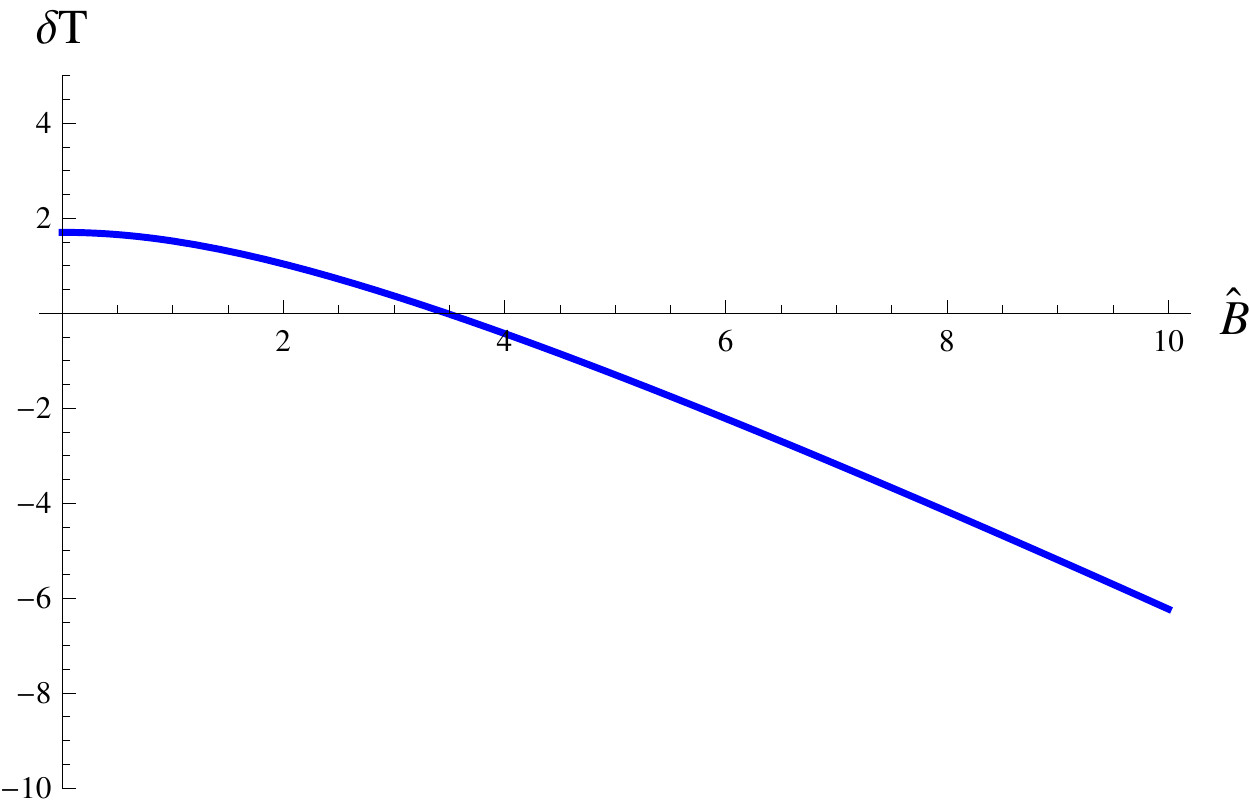}
\caption{\label{deltaTvsB} The temperature correction $\delta T$ in units of $(M_{KK}\lambda^2)/(216 \pi^4 N_c)$ as a function of the magnetic field $\hat B = |\vec{B}|/( \bar \lambda M_{KK}^2)$.  At zero magnetic field $\delta T$ is positive and then falls as $\hat B$ increases becoming negative at  $\hat B \approx 3.48$.}
\end{center}
\end{figure}

In order to solve this problem we consider a correction to the deconfinement temperature $\delta T$ of order $\lambda^2/N_c$ that would modify the gluon contribution to the pressure so that
\beqa
P_{\rm Conf}^{\rm Gluons}(T_c + \delta T) +  P_{\rm Conf}^{\rm Quarks}(T_c) =
P_{\rm Deconf}^{\rm Gluons}(T_c + \delta T) +  P_{\rm Deconf}^{\rm Quarks} (T_c) \,.
\label{TotalPressures}
\eeqa
Using the results (\ref{ConfYMPressure}), (\ref{DeConfYMPressure}), (\ref{ConfQuarkPressure}) and (\ref{DeconfQuarkPressure}) in (\ref{TotalPressures}) we find
\beqa
\delta T &=& \left ( \frac{M_{KK}}{216 \pi^4} \right ) \frac{\lambda^2}{N_c}
\Big \{ - B(-7/6,1/2) \,   F \left ( - \frac12, - \frac76 ; - \frac23 ; - \frac{ \hat B^2}{4} \right)   \cr
&-&  \frac67 \,  F \left ( - \frac12, - \frac76 ; - \frac16 ; - \frac{ \hat B^2}{4 } \right ) \Big \} \,.
\eeqa
We plot in Figure \ref{deltaTvsB} the temperature correction $\delta T$, in units of $\lambda^2/(216 \pi^4 N_c)$, as a function of the effective magnetic field $\hat B$. The main feature of this plot is that $\delta T$ decreases as the magnetic field increases, implying a lower deconfinement temperature in the presence of a strong magnetic field. This  is in accordance with the results obtained from  lattice QCD \cite{Bali:2011qj}, the MIT bag calculation \cite{Fraga:2012fs} and large-$N_c$ counting rules \cite{Fraga:2012ev}. Note, however, that at large magnetic fields $\delta T$  has a linear dependence in $\hat B$, in contrast to the saturation phenomena observed in lattice results \cite{Bali:2011qj}.

Since we are working in the antipodal scenario, the deconfinement temperature coincides with the chiral restoration temperature. Then we conclude from our results that, in the antipodal Sakai-Sugimoto model, a strong magnetic field would tend to inhibit confinement and chiral symmetry breaking. This is in contrast with the phenomena of magnetic catalysis where a strong magnetic field enhances chiral symmetry breaking, which was obtained in the non-antipodal case \cite{Bergman:2008sg,Johnson:2008vna,Callebaut:2013ria}.

\subsection{Total pressure and perturbative backreaction analysis}

The total pressure for the confined  phase can be written as
\beqa
P_{\rm Conf}^{\rm Gluons} + P_{\rm Conf}^{\rm Quarks}
&=& \frac{2}{81 \pi} M_{KK}^4 \bar \lambda N_c^2
\Big \{ 1  - \bar \epsilon \, \frac{81}{2\pi}  B(-7/6,1/2) \cr 
&\times&  F \left ( - \frac12, - \frac76 ; - \frac23 ; - \frac{ \hat B^2}{4} \right)  \Big \} \, , \label{totPConf}
\eeqa
where $\bar \epsilon := \bar \lambda^2/N_c$. Similarly, the total pressure for the deconfined phase can be written as
\beqa
P_{\rm Deconf}^{\rm Gluons} + P_{\rm Deconf}^{\rm Quarks}
&=& \frac{2}{81 \pi} M_{KK}^4 \bar \lambda N_c^2 \bar T^6
\Big \{ 1  + \bar \epsilon \, \bar T \,  \frac{81}{2\pi} \left ( \frac67 \right ) \cr
&\times&  F(-1/2,-7/6 ;-1/6 ;- \frac{ \hat B^2}{4 })  \Big \} \, , \label{totPDeconf}
\eeqa
where $\bar T =  2 \pi T/(M_{KK})$. We see from (\ref{totPConf}) and (\ref{totPDeconf}) that, although we have limited to the quenched approximation ($N_f \ll N_c$), we have already obtained a first correction to the on-shell action that is proportional to $\bar \epsilon = \bar \lambda^2/N_c$.

The perturbative parameter $\epsilon$ appears already when considering the ratio of the effective couplings for the quark and gluon actions \cite{Burrington:2007qd} :
\beqa
\tilde \epsilon := \frac{\left( \frac{\mu_8}{g_s} \right )}{ \left ( \frac {1}{16 \pi G_{10} \,  g_s^2} \right )} = \left (\frac{1}{8 \pi^2 M_{KK}^2 R^3} \right ) \frac{\lambda^2 }{N_c}  \sim \bar \epsilon \,.
\eeqa

As shown in \cite{Burrington:2007qd}, the backreaction of the quark sector can be investigated perturbatively in $\tilde \epsilon$. The result was that at linear order in $\tilde \epsilon$ only the metric and the dilaton suffer a correction, i.e.
\beqa
g^0_{PQ} &\to& g_{PQ}^0 + \tilde \epsilon \, g_{PQ}^1 \, \cr
\phi^0 &\to& \phi_0 + \tilde \epsilon \, \phi^1 \,,
\eeqa
where $g^0_{PQ}$ and $\phi^0$ are the metric and dilaton corresponding to the D4-brane background (\ref{confinedbackground},\ref{deconfinedbackground}) where as  $g^1_{PQ}$ and $\phi^1$ denote the first corrections to the metric and dilaton respectively.
Expanding at linear order in $\tilde \epsilon$ the supergravity action, that describes the gluon sector, we obtain
\beqa
S^{\rm Sugra} (g^0_{PQ} + \tilde \epsilon g^1_{PQ} , \phi^0 + \tilde \epsilon \phi^1) &\approx&
 S^{\rm Sugra} (g^0_{PQ} , \phi^0) + \tilde \epsilon \int d^{10} x
 \Big [ \frac{ \delta S^{\rm Sugra}}{ \delta g^0_{PQ}}  g^1_{PQ} \cr
 &+&   \frac{ \delta S^{\rm Sugra}}{ \delta \phi^0} \phi^1 \Big ] \,.
\eeqa
Since $g^0_{PQ}$ and $\phi^0$ are solutions to the supergravity equations we have
\beqa
\frac{\delta S^{\rm Sugra}}{ \delta g^0_{PQ}} = \frac{ \delta S^{\rm Sugra}}{ \delta \phi^0} = 0 \,.
\eeqa
From this result we conclude that the linear corrections to the metric and dilaton would not modify the on-shell supergravity action at linear order in $\tilde \epsilon$. As a consequence, the pressures obtained  in (\ref{totPConf}) and (\ref{totPDeconf}) would not receive corrections from the backreacted metric at linear order in the perturbative paremeter\footnote{An explicit example where the unquenched and quenched result for the pressure coincide at linear order was obtained in \cite{Bigazzi:2011db,Mateos:2007vn} for the D3-D7 brane model.}.

\section{Conclusions}

We have investigated the thermodynamics of massless quarks in the presence of a magnetic field considering a holographic model for large-$N_c$ QCD. From the renormalized euclidean actions for the probe branes we have estimated the quark contribution to the pressure and the magnetic susceptibilities. We found positive susceptibilities indicating a paramagnetic behaviour and we estimated a $\lambda^2/N_c$ correction to the deconfinement temperature that decreases as a function of the magnetic field. This result can be interpreted as a magnetic inhibition of confinement and chiral symmetry breaking. It is important to mention that a similar phenomena called {\it inverse magnetic catalysis}, which is the magnetic inhibition of chiral symmetry breaking, has been recently found in the Sakai-Sugimoto model when a chemical potential is turned on \cite{Preis:2010cq}.

We have restricted to the antipodal limit of the Sakai-Sugimoto model. It would be interesting to extend this work to the non-antipodal case where chiral symmetry restoration can occur after the deconfinement transition. For simplicity, we have renormalized the DBI actions considering a minimal subtraction procedure based on the principle of least action. A more formal subtraction can be done using covariant counterterms but since the asymptotics is not $AdS$ it might be necessary to map the D4-brane background to an asymptotically $AdS_6 \times S^4$ spacetime, as suggested in \cite{Kanitscheider:2008kd,Kanitscheider:2009as,Benincasa:2009ze}. We leave these issues for a future work.

It is important to remark that we have worked in the probe approximation so we did not  take into account the backreaction of the D8-$\overline{{\rm D}8}$ branes on the D4-brane background. Nevertheless, we conjectured that the deformed D4-brane background would contribute only at quadratic order in the perturbative parameter so that our result in the probe approximation would not be modified. To check this we have to evaluate the supergravity action for the deformed background using the methods developed in \cite{Burrington:2007qd}. As a bonus we may find quadratic corrections to the pressures.

\section*{Acknowledgements}

I would like to thank Kasper Peeters, Marija Zamaklar, Gast\~ao Krein, Jorge Noronha, Eduardo Fraga, Francisco Rojas and Gergely Endrodi for very helpful discussions. This work is supported by Capes (Brazilian fellowship).

\begingroup\raggedright\endgroup

\end{document}